\documentstyle[prl,aps,preprint]{revtex}
\begin{document}
\draft
\title{Two routes to metallic behavior for a  Kondo insulator} 
\author{A. Schr\"{o}der$^{1}$, G. Aeppli$^{1,2}$, T. E. Mason$^{1,3}$, 
E. Bucher$^{4,5}$, C. Broholm$^{6}$, K.N. Clausen$^{1}$\\}
\address{$^{1}$ Ris{\o} National Laboratory, 4000 Roskilde, Denmark}
\address{$^{2}$ NEC, 4 Independence Way, Princeton, NJ 08540, U.S.A.}
\address{$^{3}$ University of Toronto, Dept. of Physics, Toronto, M5S 1A7, Canada}
\address{$^{4}$ University of Konstanz, Konstanz, Germany} 
\address{$^{5}$ Bell Laboratories, Lucent Technologies, Murray Hill, NJ 07974, 
U.S.A.}
\address{$^{6}$ The Johns Hopkins University, Baltimore, Maryland 21218, U.S.A.}
\date{Submitted to Physical Review Letters, November 18, 1996}
\maketitle
\begin{abstract}
Cold neutron spectroscopy was performed on CeNi$_{1-x}$Cu$_x$Sn to investigate 
how the magnetic gap in the 'Kondo insulator' CeNiSn is destroyed by 
temperature ($T$), chemical substitution ($x$), and external magnetic field. 
Upon doping, the spin gap collapses and magnetic Bragg peaks occur initially 
($x$ = 0.13) at the commensurate wave vectors $Q$ = (0,m/2,l) (m,l integer) 
where the higher-energy (4.1 meV) gap is located for $x$ = 0. A magnetic field 
smooths the sharp spin gap structure in both momentum and energy, while 
leaving the associated static susceptibility unchanged, as does raising T. 
\end{abstract}


It is by now widely appreciated that antiferromagnetic fluctuations
are a key feature of strongly interacting Fermi systems, be they transition
metal oxides or rare earth intermetallics. Such fluctuations
play an  important role and are indeed drastically modified near the 
metal-insulator transition of the oxides \cite{1}. Transitions of this 
type are generally referred to as Mott-Hubbard transitions, and are driven by 
the immobilization of carriers due to Coulomb repulsion. One outstanding 
question is whether antiferromagnetic fluctuations play an equally important 
role near the metal-insulator transition
in intermetallic compounds, which, while they may be close to being band
insulators, have strongly renormalized bands due to Coulomb interactions. In
these compounds, also called Kondo insulators \cite{2,3}, there are several 
routes to metallic behavior.
The first (Fig.\ \ref{fig1}(b)) is chemical doping, as in the most celebrated of 
semiconductors, Si. The second (Fig.\ \ref{fig1}(c)), not feasible for
Si because of its large band gap, but feasible for the small gap Kondo
insulators, is to apply a magnetic field for which the Zeeman energy is of 
order the band gap. 
A Fermi surface emerges because there are no longer two
spin states  below the Fermi level for each momentum index. 
In the present paper, we show that these two routes to metallicity are
very different in CeNiSn, the most studied Kondo insulator 
\cite{4,5,6,7,8,9,10,11,band,12}. CeNiSn
is considered a Kondo or strongly correlated insulator because it is a
band insulator with measured gaps \cite{4,5,6,7,8,9,10,11} to spin and
charge excitations, which are much smaller than those predicted by
band theory \cite{band}. More specifically, there is an extraordinary loss of
carriers at low temperatures, even though the $<$1/200 carriers/unit cell which
remain belong to a conduction sea characterized by a mean free path 
which is steadily increasing with improved  sample preparation \cite{7}.
With Cu doping onto the Ni site, antiferromagnetic interactions
emerge which are sufficiently strong to yield an antiferromagnetic (AFM) ground
state \cite{4,5}; the phase diagram (Fig.\ \ref{fig1}(d)) is the
reverse of the standard phase diagram for the transition metal oxides where
doping transforms an AFM insulator into a peculiar paramagnetic metal.
We report here the counter-intuitive discovery that 
at low doping the order occurs at the commensurate wave vector (which is 
unusual for Ce-based heavy fermion AFM) $Q_0$ for 
which the spin gap in the parent compound is of highest energy.  On the other 
hand, application of a magnetic field $H$ appears mainly to smooth out the 
sharp momentum and energy-dependent features  characteristic of $H$ = 0 and
does not affect the screening of the AFM interactions at $\omega = 0$. 

Our samples are the pure CeNiSn crystals used in \cite{9},
a Czochralski-grown $x$ = 0.13 boule (the composition was established
by high-resolution neutron diffraction and the known linear relation
between lattice constants and $x$ for the orthorhombic ($\epsilon$-TiNiSi) 
CeNi$_{1-x}$Cu$_x$Sn series \cite{4,13}) containing a main crystal with a 2$^o$ 
mosaic, and an $x$ = 0.3 arc-melted polycrystalline ingot. We 
employed primarily the RITA \cite{14} instrument at TAS7 installed in the Ris\o 
\hspace{0.1cm}cold neutron guide hall. For most scans shown below, a flat 
pyrolitic graphite (PG) analyzer collected $E_f$ = 3.5 meV scattered neutrons;
the incoherent resolution here has a full-width-at-half-maximum (FWHM) of 0.17 
meV. To increase the intensity for the much smaller $x$ = 0.13 sample we 
installed the RITA 7-blade horizontally focussing PG-analyzer for 
E$_f$ = 5.5 meV (with resolution FWHM = 0.33 meV). 
Magnetic fields were applied perpendicular to the [100] scattering plane. 

The principal features in the magnetic excitation spectrum of CeNiSn are peaks 
with sharp leading edges and long tails to high energies found at special 
momentum transfers $Q$. Such peaks exist at an energy transfer of 
$\hbar\omega_{0}$ = 2.5 meV for momenta equivalent
to (0,0,1) \cite{9}, and at $\hbar\omega_{0}$ = 4.1 meV for momenta such as 
$Q_0$=(0,0.5,1) \cite{10}. Figure \ref{fig2}(a) shows the
latter at various $T$ for $x$ = 0 and $H$ = 0. Warming to $T$ = 15 K, a 
temperature well below the gap value $\hbar\omega_{0}/k_B$, results in very 
substantial broadening of the peak. Thus, the emergence of the high 
energy peak as a sharp feature coincides with the onset of 
low $T$ coherence and the loss of carriers identified in the bulk 
transport measurements on CeNiSn \cite{7}. 
Figure \ref{fig3}(a) shows a constant-$E$ scan for an energy
transfer fixed at the peak position, 4.1 meV. It is apparent that the
4.1 meV feature is also sharp in $Q$-space, with its maximum at 
$Q_0$ = (0,0.5,1). The corresponding magnetic coherence length along [0,1,0], 
associated with the Lorentzian, ($\sim 1/((Q-Q_0)^2 \xi^2+1))$, 
which is drawn as a solid line through the data, is $\xi$ = 24 $\pm$ 4 \AA. 
Thus, what we are seeing here is not very
different from the various finite frequency resonance and threshold
phenomenon found  in high-$T_c$ superconductors below their
transition temperatures \cite{15}. The superficial similarity is not so 
surprising given that both superconductivity and the onset of coherent 
semimetallic or insulating behavior entail the replacement of 
ungapped excitation continua by spectra with vanishing or at least
reduced weight at low frequencies and sharp threshold effects at higher
energies. Indeed, we find  that the 4.1 meV resonance can be described 
using the same BCS-inspired form used previously \cite{9} to account for the
2.5 meV resonance,
\begin{equation}
\frac{S(Q,\omega)}{n(\omega)+1} = \chi''(Q,\omega) =  \chi_{0} \mid\epsilon
\mid \omega\: (Re(\frac{1}{\sqrt{\omega^{2}-\epsilon^{2}}}))^{2}, 
\end{equation}
with $\epsilon = \Delta/\hbar + i\Gamma$,
where $\Delta/\hbar$ and $\Gamma$ represent the frequency and decay rate of the 
resonance. Even so, further investigation of CeNiSn reveals an additional 
feature of a type not uncovered for superconductors. Specifically 
Fig.\ \ref{fig3}(b) demonstrates that at $\hbar\omega$ = 3 meV, there is 
actually a minimum in the magnetic scattering at $Q_0$ = (0,0.5,1), where the 
maximum existed for $\hbar\omega_0$ = 4.1 meV. The sharp minimum for 
$\hbar\omega$ = 3 meV nearly compensates for the sharp maximum for 4.1 meV  
in the sense that only a weak $Q$-dependent modulation remains when the two 
scans are added together with the 1/$\omega$ weights needed to compute the 
real part of the susceptibility $\chi'(Q,\omega$=0) from $\chi''(Q,\omega)$ 
(see Fig.\ \ref{fig3}(c)) using the Kramers-Kronig formula,
\begin{equation}
\chi'(Q,\omega=0) =1/\pi P \int_{-\infty}^{\infty}
\chi''(Q,\omega)/\omega \: d\omega.
\end{equation}
Thus, like the 2.5 meV resonance \cite{9}, the 4.1 meV resonance \cite{10} 
is associated with a strongly $Q$-dependent $\chi''(Q,\omega)$, but only a 
weakly $Q$-dependent $\chi'(Q,\omega$=0). Also, as was found for $Q$ near 
(0,0,1) \cite{9}, where the 2.5 meV resonance dominates, there is essentially 
no $T$ dependence of $\chi'(Q_0,\omega$=0) obtained from the full 
Kramers-Kronig transform \cite{16} of energy scans such as those shown in
Fig.\ \ref{fig2}(a). This $T$-independence, illustrated in 
Fig.\ \ref{fig4}(a), occurs notwithstanding the strong $T$-dependence of 
$\chi''(Q_0,4.1$meV) shown in Fig.\ \ref{fig4}(b).

We turn now to how two different perturbations, both of which lead to more 
metallic behavior, have radically
different consequences for both $\chi'(Q,\omega$=0) and $\chi''(Q,\omega)$.
Figure \ref{fig2}(b)  shows the effect of an external magnetic field applied 
along the $a$-axis, which is the direction for which the bulk susceptibility 
is known to be highest \cite{6}. There is a very clear reduction and 
broadening of the 4.1 meV resonance at low $T$. 
In addition, as Fig.\ \ref{fig3}(d) demonstrates, broadening
occurs in the $Q$- as well as the frequency dependence of the 4.1 meV 
feature. Furthermore, the compensating minimum in the $\hbar\omega$ = 3 meV
scan disappears (Fig.\ \ref{fig3}(e)), leaving our estimate of the 
$Q$-dependent $\chi'(Q,\omega$=0) (Fig.\ \ref{fig3}(f)) essentially unchanged. 
The same applies to $\chi'(Q_0,\omega$=0), as calculated more rigorously from 
the constant-$Q_0$ scans and Eq.(2). Indeed, Figs 4(a) and (c) show
that the resulting $T$- and $H$-dependent values are indistinguishable from 
those for $H$ = 0, even though the field attenuates the $T$-dependence of 
the peak maximum $\chi''(Q_0,\hbar\omega_0$=4.1meV)
(Fig.\ \ref{fig4}(b)) and does of course reduce $\chi''(Q_0,\omega_0$) very 
noticeably at low $T$ (Fig.\ \ref{fig4}(d)).

Our observation of the $H$-dependent $\chi''(Q,\omega)$ can be modelled 
using Eq.(1) in at least three ways. The first assumes that raising  
$H$ merely increases $\Gamma$ for fixed $\Delta$. The second and third 
invoke a Zeeman splitting $\delta = g\mu_BH$ of the excited state 
associated with the 4.1 meV peak, assuming that it is either a doublet
or a triplet with a lifetime $\Gamma^{-1}$ unaffected by $H$.
The band picture of Fig.\ \ref{fig1} gives rise to the triplet model as long as 
$g\mu_BH$ is below the initial band splitting. The lines in 
Figs.\ \ref{fig2}(b) and \ref{fig4}
correspond to fits using the three models. The outcomes do not differ
significantly, although there is a slight preference, most noticeable in
Fig.\ \ref{fig4}(d) at low $H$, for the triplet model using the eminently
sensible value of $g\mu_B = 2 \mu_B$.
The conclusion is that while an external magnetic field 
greatly smooths the 4.1 meV resonance which characterizes the
low-carrier density coherent state of CeNiSn, it does not
noticeably affect the zero frequency interactions. Somehow, the
perturbation due to the external field, which does not alter 
the number of electrons in the 
problem, also does not affect the screening responsible
for the essentially $Q$- and $T$-independent $\chi'(\omega$=0). 

We next show that chemical doping, which changes the number of electrons 
\cite{8}, destroys the screening. In particular, we have discovered that
at low $T$, sufficiently doped CeNi$_{1-x}$Cu$_x$Sn orders 
magnetically with wave vectors close to reciprocal space points of the 
form (0,m/2,l) with m,l integer, the locations
of the higher energy (4.1 meV rather than 2.5 meV) resonances in
pure CeNiSn. For $x$ = 0.13, the peaks occur below $T_N$ = 1.4 K precisely 
at the commensurate half-order points, as shown for example in
the scan through a resolution-limited magnetic superlattice peak plotted 
in the inset of Fig.\ \ref{fig5}(a). On the other hand, for our more heavily 
doped ($x$ = 0.3) sample, the peaks appear incommensurate, with the lowest
ordering vector, with $|Q|$ = 0.72 \AA$^{-1}$, slightly larger than 
$|Q_0|$ = 0.69 \AA$^{-1}$. Fig 5(a) shows the $T$-dependences for 
representative magnetic Bragg peaks in the doped materials. The ordering 
temperatures $T_N$ are not far from resistance anomalies reported for the 
same compound and indeed used to construct the magnetic phase diagram
of Fig.\ \ref{fig1}(d) for CeNi$_{1-x}$Cu$_x$Sn \cite{4}.
Because the doped material orders magnetically via a second-order
phase transition, $\chi'(Q)$ must display a non-trivial $Q$- and 
$T$-dependence. Figure \ref{fig2}(c), which shows the frequency-dependent 
scattering at the ordering vector $Q_0$ = (0,0.5,1), and Fig.\ \ref{fig5}(b), 
which shows the dramatic rise with cooling of the low-$\omega$ 
$\chi''(Q_0,\omega)$, allows us to see directly how these attributes 
develop. All hints of the resonance in Figs.\ \ref{fig2}(a) and
(b) have disappeared, and the quasielastic form 
$S(Q,\omega)=(n(\omega)+1) \chi_0\omega\Gamma/(\Gamma^2+\omega^2)$ 
describes the data well. As for the pure compound, we can use the 
Kramers-Kronig relation to extract $\chi'(Q_0)$, which rises rapidly with 
decreasing $T$ (see Fig.\ \ref{fig5}(c)) to a value very much 
larger than the $T$-independent value for $x$ = 0.  

In conclusion, we have shown that two approaches to conventional
heavy fermion behavior in the 'Kondo insulator' CeNiSn yield
radically different effective interactions between magnetic moments. 
A 9T magnetic field, which conserves electron density,
appears to broaden the dominant finite frequency resonance 
(left inset in Fig.\ \ref{fig1}(d)) in both
frequency and momentum spaces, but does not appear to shift it.
Furthermore, it does not affect the screening of the
zero-frequency interactions which is one of the Hallmarks of
compensated 'Kondo insulators'. Chemical doping, which alters
the electron density, yields dramatically different results, in that
it eliminates both the resonances and the screening mechanism to produce
a relatively conventional heavy fermion antiferromagnet 
(right inset in Fig.\ \ref{fig1}(d)). Our 
experiments therefore constitute strong evidence for the critical
role of carrier density in the heavy fermion problem. Thus, heavy fermion
compounds inhabit a three-dimensional space labelled by carrier density
in addition to the familiar Kondo and RKKY couplings whose competition
has historically been held accountable for the peculiarities of these 
fascinating materials.

We thank T. Takabatake, H. Kadowaki, S. Raymond, A. S. Mishchenko and 
C. M. Varma for helpful discussions.

\newpage
\begin{figure}
\caption[schematic]{(a-c) Schematic band structure (energy $\epsilon$ vs momentum $Q$) 
of a Kondo 
insulator and related metallic states. The starting point is a half-filled
(n=1/2) dispersive conduction band coupled to a half-filled dispersion-less
band containing more localized f-electrons. A hybridization gap ensues, and
an insulating state obtains because the band below the gap is fully occupied
by states of mixed conduction- and f-like character. Metal-insulator 
transitions are achieved by moving the Fermi level by chemical doping (b), or
by using an external-field-induced Zeeman splitting to sweep the bands through 
the Fermi level (c).
(d) $T_N$, derived from resistivity data \cite{4}, vs $x$ for 
CeNi$_{1-x}$Cu$_x$Sn. Squares correspond to the onset of magnetic order from 
neutron diffraction. The insets represent schematically the principal loci, 
in the momentum-energy transfer plane, of the magnetic spectral function for 
undoped and doped CeNiSn: doping causes a sharp  spin-gap-like feature in
the paramagnetic Kondo insulator (PM) to collapse into a zero-frequency 
Bragg peak in the heavy fermion antiferromagnet (AFM).}
\label{fig1}
\end{figure}

\begin{figure}
\caption{Constant-$Q$ scans for CeNi$_{1-x}$Cu$_x$Sn at wave vector 
$Q_0$ = (0,0.5,1) for low (closed) and high temperatures $T$ (open symbols).
The background-corrected neutron intensity is shown for $x$ = 0 in zero 
field $H$ = 0 (a) and in high magnetic field $H$ = 8.9 T (b) applied along the 
$a$-direction and for the doped compound with $x$ = 0.13 (c).
Solid lines represent fits to the gap function of Eq.(1), thick and broken
lines are for the models described in the text.}
\label{fig2}
\end{figure}

\begin{figure}
\caption{Constant-$E$ scans for CeNiSn along $Q$ = (0,k,1) for $H$ = 0 (a-c) 
and for $H$ = 8.9 T (d-f) at the peak frequency (a,d) and below (b,e).
(c,f) show $\Sigma \chi''/\omega \sim$ I(4.1meV)/4.1meV + I(3.0meV)/3.0meV, 
which provides an estimate of $\chi'(\omega$=0). Lines are Lorentzians, 
superposed on flat backgrounds.}
\label{fig3}
\end{figure}

\begin{figure}
\caption{$\chi''(Q_0,\omega_0$) at resonance vs. temperature $T$
(b) in $H$ = 0 and $H$ = 8.9 T and vs. magnetic field $H$ (d) at low $T$.
As in Fig.\ \ref{fig2}(b), the lines correspond to the different models
described in the text.
Triangles show the fitted Zeeman splitting value $\delta$ of the triplet model
with constant $\Gamma(H)$, yielding $g\mu_B = 2\mu_B$.
The upper panel displays the T- (a) and H-independence (c) of 
$\chi'(Q_0,\omega$=0) derived from Eq.(2).}
\label{fig4}
\end{figure}

\begin{figure}
\caption{(a) Intensity (normalized to yield unity at the lowest measuring
temperature in each case) at the ordering vectors $Q_0$ = (0,0.5,0) 
for $x$ = 0.13 and $Q_{IC}$ = 0.72 \AA$^{-1}$ for $x$ = 0.3 vs. temperature. 
Inset shows the magnetic Bragg profile at $Q_0$ for $x$ = 0.13.
(b) Low frequency magnetic response $\chi''$(Q$_0,\hbar\omega$=0.75meV) 
at $Q_0$ = (0,0.5,1) for $x$ = 0.13 and $x$ = 0.
(c) $\chi'(Q_0,\omega$=0) derived from Eq.(2) for $x$ = 0.13 and $x$ = 0. 
Vertical scales for (b) and (c) and the two compositions are the same, 
having been normalized using acoustic phonons near (0,2,0). The values 
shown for $x$ = 0.13 are divided by 3. Lines are guides to the eye.}
\label{fig5}
\end{figure}


\begin{references}
\bibitem{16} 
$\chi$' was evaluated from the fit function (Eq.(1)) of $\chi''(\omega)$,
and does not differ from the Kramers-Kronig sum over the experimental spectra
in the restricted energy range, up to an additive constant to do with
 the unobserved high-frequency response, which is  most likely
(given the available data) $T$- and $H$-independent.
\end{references}
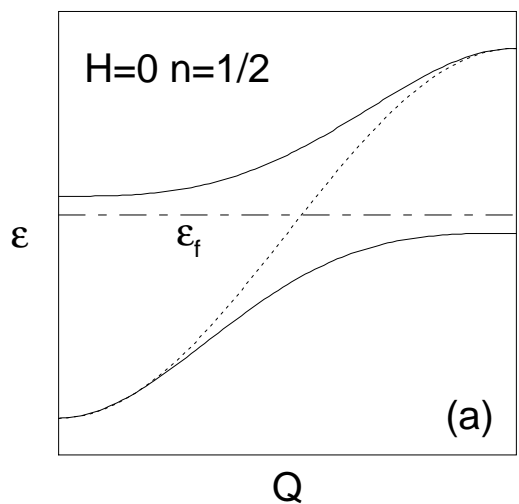
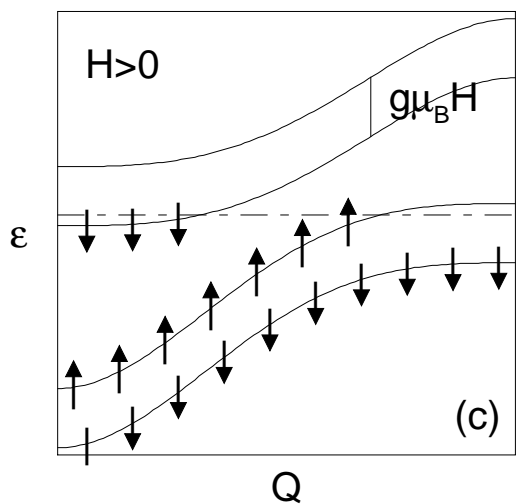
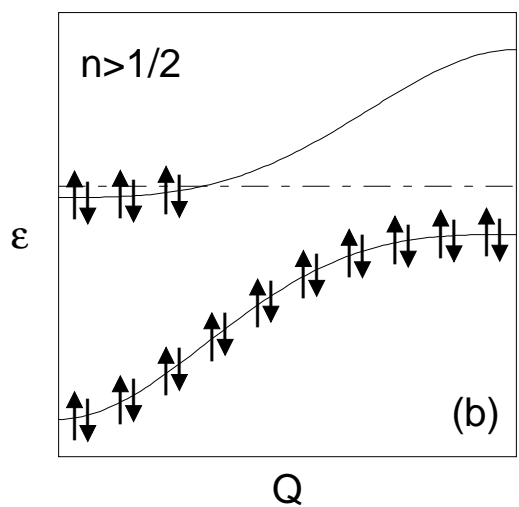
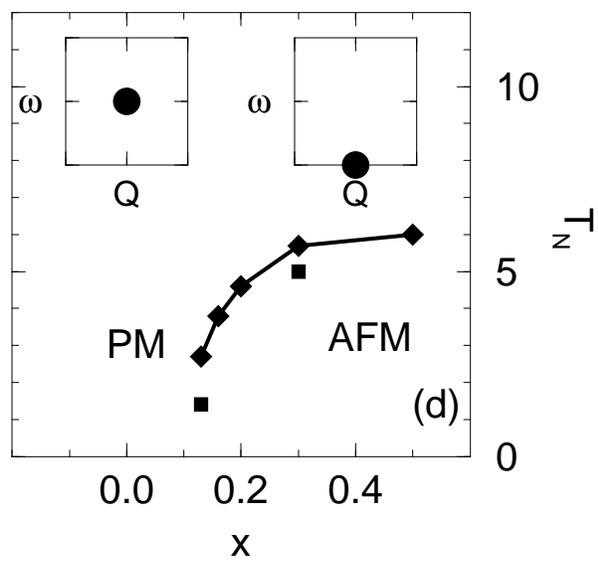
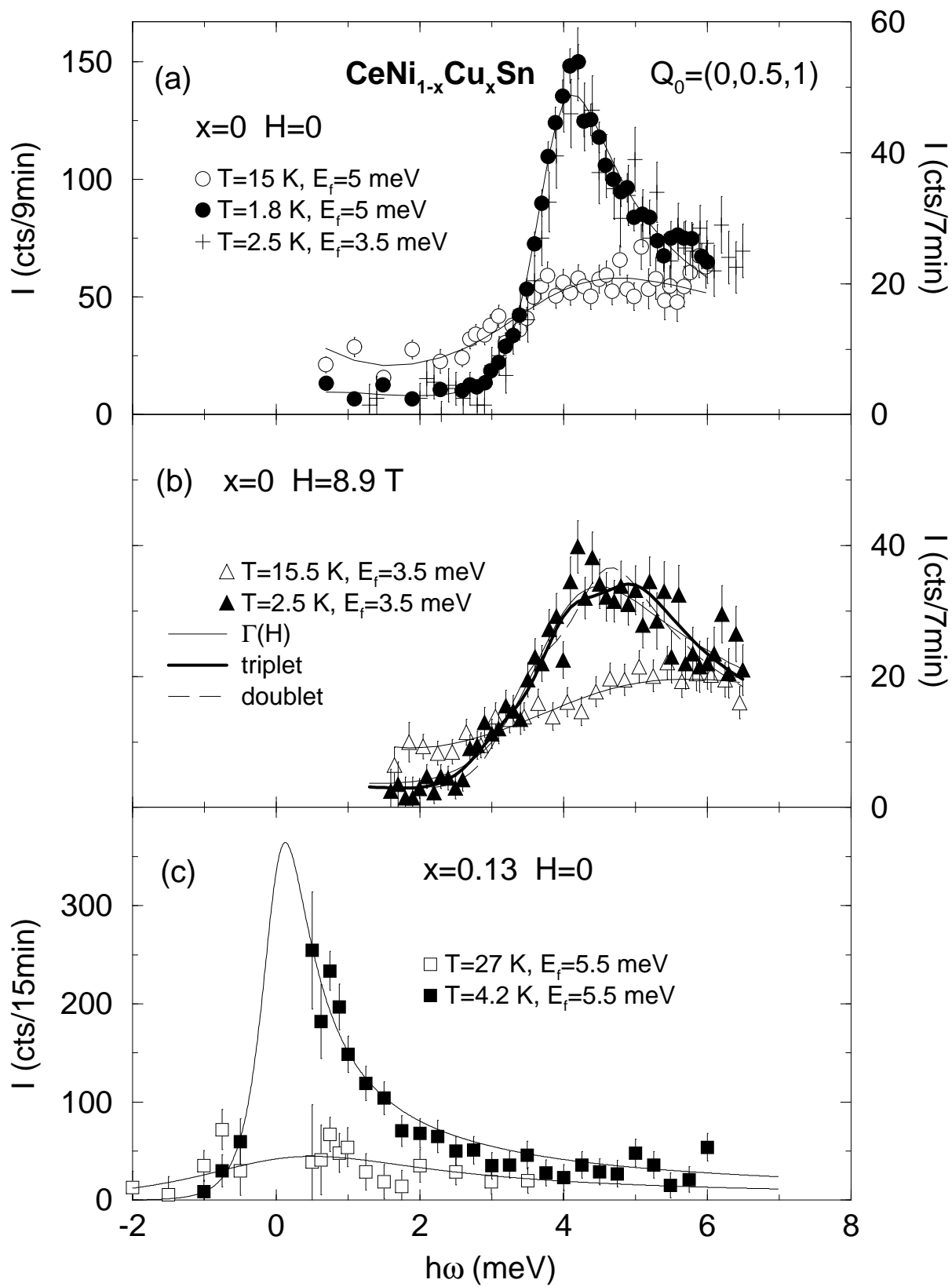
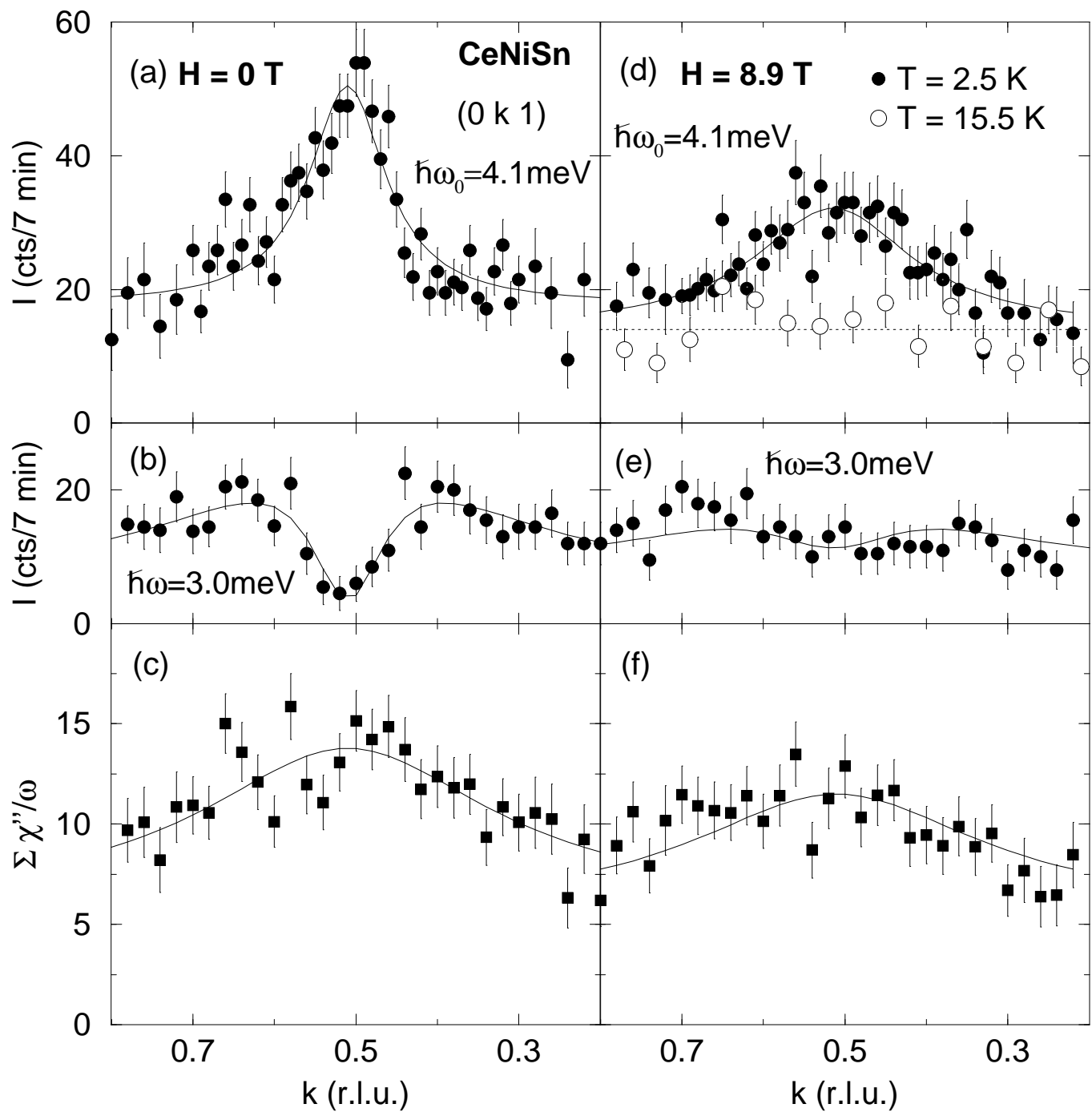
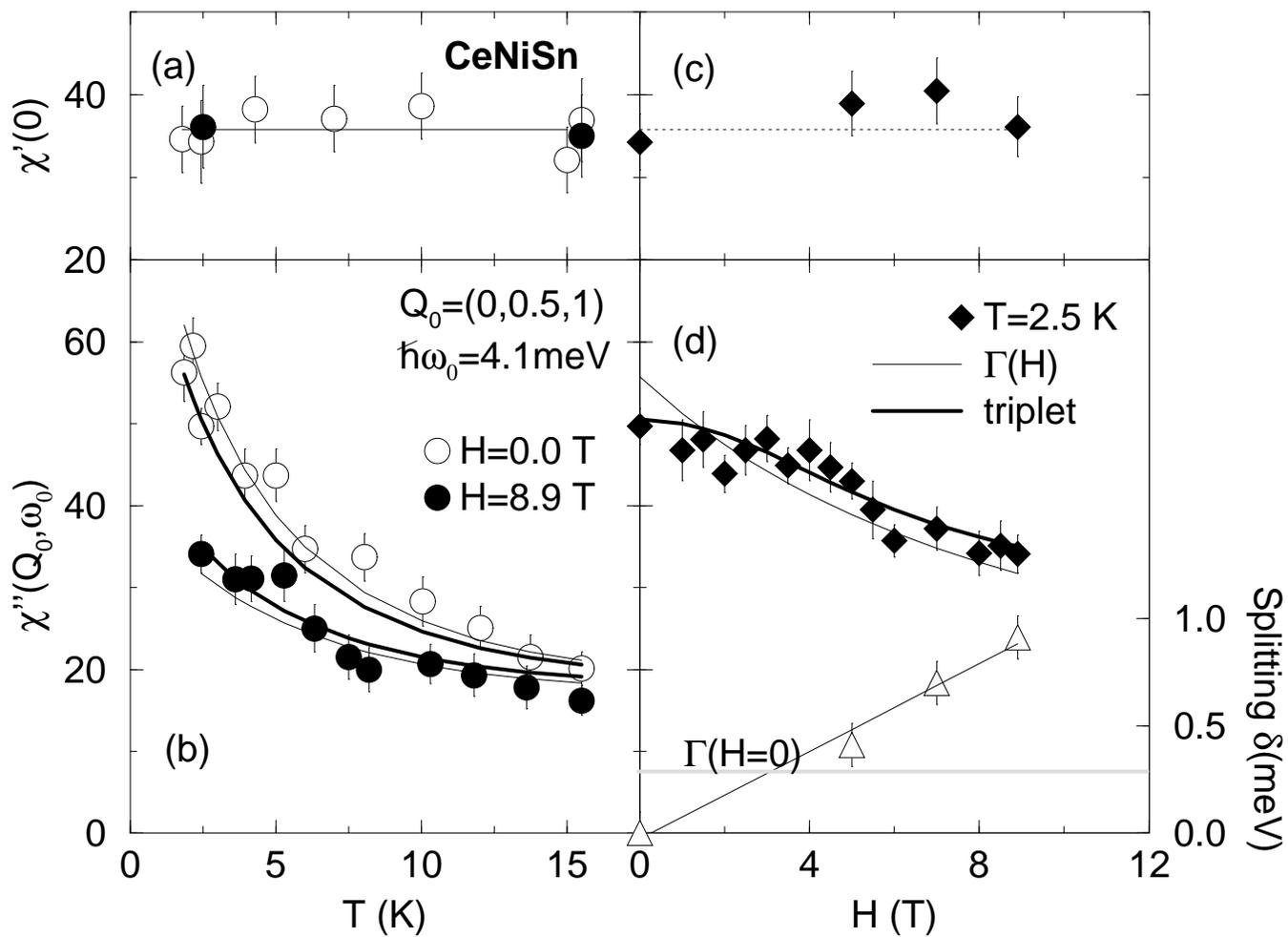
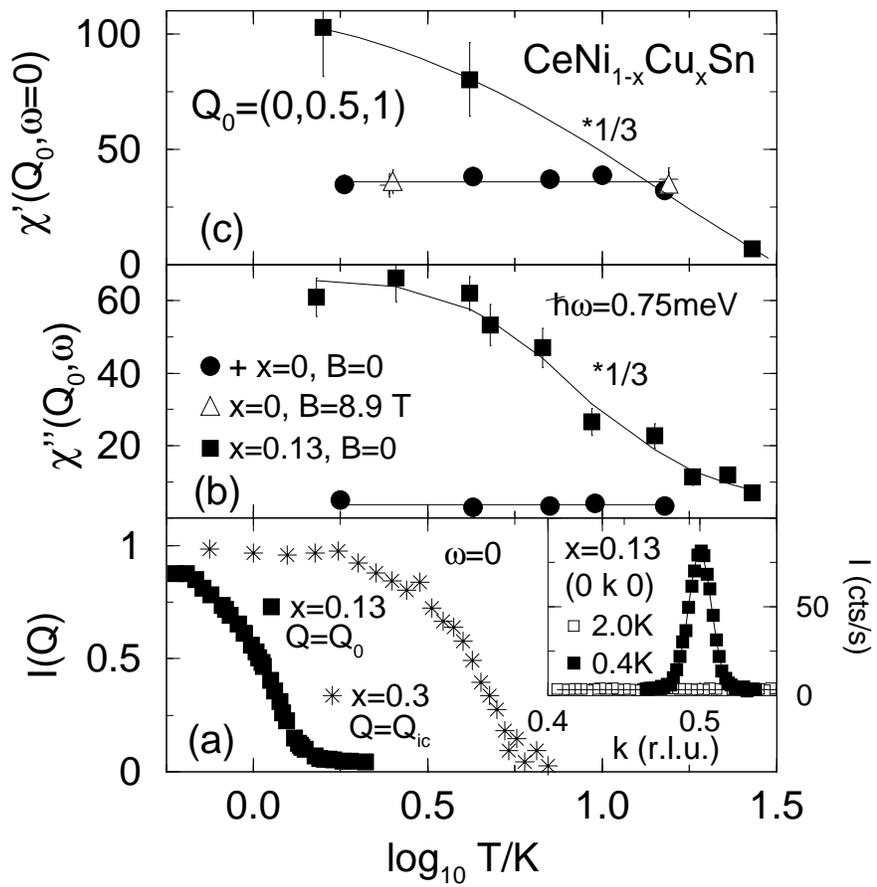
\end{document}